\documentclass[a4paper]{aa}
\usepackage{psfig,longtable,lscape}
\usepackage{epsfig}
\usepackage{graphicx}
\def \xmm {XMM--Newton}
\def \sax {BeppoSAX}
\def \src {LS~I~+61~303}

\def \hcm {\hbox {\ifmmode $ atom cm$^{-2}\else atom cm$^{-2}$\fi}}
\def \arcmin {\hbox{$^\prime$}}
\def \arcsec {\hbox{$^{\prime\prime}$}}

\newcommand{\loe}{\stackrel{<}{\sim}}
\newcommand{\goe}{\stackrel{>}{\sim}}

\begin{document}
   \title{\xmm\ observation of a spectral state transition in the peculiar radio/X--ray/$\gamma$--ray source \src 
\thanks{Based on obervations obtained with XMM-Newton,
an ESA science mission with instruments and contributions directly funded by ESA Member States and NASA.}}


   \author{L.Sidoli\inst{1}
          \and
	  A.Pellizzoni\inst{1}
          \and
          S.Vercellone\inst{1}
          \and
	  M.Moroni\inst{1}
          \and
          S.Mereghetti\inst{1}
          \and 
          M.Tavani\inst{2,3,4}
        }

   \offprints{L.Sidoli (sidoli@iasf-milano.inaf.it)}

   \institute{Istituto di Astrofisica Spaziale e Fisica Cosmica --
        Sezione di Milano -- INAF-IASF, Milano, I-20133, Italy\\
              \and
CIFS, Torino, I-10133, Italy
\and
        Istituto di Astrofisica Spaziale e Fisica Cosmica --
        Sezione di Roma -- INAF-IASF, Roma, I-00133, Italy
\and
Dipartimento di Fisica, Universit\`a di Tor Vergata, Roma, I-00133, Italy
             }

   \date{Received  29 June, 2006; Accepted 24 August, 2006}

\authorrunning{L. Sidoli et al.}

\titlerunning{{X--ray spectral state transition in \src}}

\abstract
{We report the results of  XMM-Newton and BeppoSAX observations of the 
radio and X--ray emitting star \src, likely associated with the gamma-ray 
source 2CG 135+01 and recently detected also at TeV energies. The data 
include a long XMM-Newton pointing carried out in January 2005, 
which provides the 
deepest look ever obtained for this object in the 0.3--12 keV range.  During 
this observation the source flux decreased from a high level of 
$\sim$13$\times$10$^{-12}$~erg~cm$^{-2}$~s$^{-1}$ to 
4$\times$10$^{-12}$~erg~cm$^{-2}$~s$^{-1}$ within 2-3 hours.
This flux range is the same seen in shorter and less sensitive 
observations carried out in the past, but the new data show for the first 
time that transitions between the two levels can occur on short time 
scales.
The flux decrease was accompanied by a significant softening of the 
spectrum, which is well described by a power law with photon index 
changing from 1.62$\pm{0.01}$ to 1.83$\pm{0.01}$.
A correlation between hardness and intensity is also found when comparing 
different short observations spanning almost 10 years and covering various 
orbital phases.
\src\ was  detected in the 15--70 keV range with the PDS instrument in 
one of the BeppoSAX observations, providing evidence for variability also 
in the hard X--ray range. The X--ray spectra, discussed in the context of  
multiwavelength observations, place some interesting constraints on the 
properties and location of the high-energy emitting region.
}
   \maketitle
%

\section{Introduction}

The peculiar radio source GT 0236+610, unambiguously associated 
with the B0 Ve star \src, 
is unique in its high variability and periodic
radio emission (Gregory \& Taylor 1978). 
The radio outbursts show a periodicity of about 26.5 days 
(Taylor \& Gregory 1982; Gregory et al., 1999) and a further modulation 
of both the outburst phase and outburst
peak flux with a period of $\sim$1600 days (Gregory et al. 1999), 
also displayed in the H$_{\alpha}$ emission line (Zamanovet al. 1999).

A faint X--ray counterpart (F$_{\rm 2-10 keV}$, 6$\times$10$^{-12}$~erg~cm$^{-2}$~s$^{-1}$) 
was identified by Bignami et al. (1981) with the Einstein satellite, 
and later observed with ROSAT (Goldoni \& Mereghetti 1995; Taylor et al. 1996), 
ASCA (Leahy et al. 1997), RXTE (Harrison et al. 2000; Greiner \& Rau 2001).
 
The hard power law X-ray spectrum extending up to 
10~keV without breaks (photon index of $\sim$1.7, 
absorbing column density, N$_{H}$, of $\sim$5$\times$10$^{21}$~cm$^{-2}$) 
is clearly inconsistent with low temperature plasma emission, 
thus excluding the Be-star as a main contribution to the X-ray emission.

The 26.5 days periodicity, reflecting the orbital motion of 
the system (Gregory \& Taylor 1978, Taylor \& Gregory 1982), 
has also been observed in the optical band (Hutchings \& Crampton 1981; Mendelson \& Mazeh 1989), 
in the infrared (Paredes et al. 1994), in soft X-rays (Paredes et al. 1997) 
and in the H$_{\alpha}$ emission line (Zamanov et al. 1999).

Spectral line observations of the radio source give a 
distance of 2.0$\pm{0.2}$~kpc (Frail \& Hjellming 1991), implying
an X--ray luminosity of 10$^{33}$~erg~s$^{-1}$. 
No periodic pulsations have been detected in its X--ray and radio emission. 
The low X-ray luminosity and the lack of iron line emission 
indicate that \src\ is not a classical accreting X-ray pulsar. 
Although this is not surprising in
view of the strong radio emission (never seen from an X--ray pulsar, see e.g. Fender 2001), 
the low X-ray luminosity is remarkable in a source that can only 
be explained with the presence of a compact object.

\src\ is very likely 
associated with one of the most interesting unidentified
gamma-ray sources near the plane of the Galaxy, 
the COS-B source 2CG~135+01 (Hermsen etal. 1977). 
Although the association between 2CG 135+01 and \src\ was initially weakened by the
presence of another plausible counterpart 
in the COS-B error region (the quasar QSO 0241+622), 
the reduced error box determined by EGRET (Kniffen et al. 1997) 
is  compatible only with the position of \src.

A recent detection of variable and likely periodic 
very high energy gamma-ray emission above 100~GeV has been reported
by the MAGIC collaboration (Albert et al. 2006).

The strong radio outbursts suggest the presence of a compact star, 
although there is no direct evidence for the existence of a 
neutron star (no pulsations nor X-ray bursts). 
Two main classes of models have been proposed for \src, 
involving a highly eccentric binary system with an orbital period of 26.5 days, 
composed of a compact star and the Be companion. 

The first class of models suggests that the radio outbursts are 
produced by streams of relativistic particles originating in 
episodes of super-Eddington accretion onto the compact object 
(e.g., Taylor et al., 1992). 
The X--ray luminosity of \src\ is orders of magnitude lower than the Eddington limit. 
The association with the
gamma-ray source 2CG 135+01 suggests that the bulk of 
the energy output is shifted from X--ray to $\gamma$--ray wavelengths, 
but the actual mechanism responsible for this is not well understood 
in the context of the supercritical accretion model.

The second class of models assumes that \src\ contains a non-accreting young 
rapidly rotating neutron star
(e.g., Maraschi \& Treves, 1981; Tavani, 1994). 
The radio outbursts are produced by energetic electrons accelerated in the 
shock boundary between the relativistic wind of the young pulsar 
and the wind of the Be star. 
This scenario is supported by the shock-powered emission 
observed by ASCA and CGRO near the periastron in a possibly 
similar system, the binary PSR B1259-63 (Tavani et al., 1995), 
which is composed by a radio pulsar (with a period of 47~ms) orbiting a Be star (orbital
period of 3.4~yr). 
In this case,the high-energy emission allows a sensible 
diagnostics of the shock emission region where the pulsar wind
interacts with the circumstellar material originating from the 
surface of the massive star (Tavani, 1994).
The modulation of the radio emission might be due to the time variable 
geometry of a `pulsar cavity' as a function of the orbital phase. 

A third way of producing X-rays has been suggested by Campana et al. (1995), 
as accretion onto the pulsar magnetosphere when the stellar wind ram pressure 
exceeds the pulsar wind pressure, but is not large enough to penetrate the 
pulsar magnetospheric boundary. 
This situation can occur in \src\ for reasonable values of 
the neutron star magnetic field and spin period. 

An object similar to \src\ has been discovered, 
the microquasar LS~5039 (Paredes et al. 2000, 2005) 
which is subluminous in the X--ray range (even more than \src) and, if the proposed
association with the EGRET source 3EG J1824-1514 is correct, also 
shows the same puzzling behavior,having
L$_{\gamma}$ $>$ L$_{\rm X}$. 

Therefore, \src\ and LS~5039 could be the first two examples of 
a new class of X--ray binaries with powerful
${\gamma}$-ray emission (see also Aharonian et al. 2005 for a HESS detection
of LS~5039 at energies above 250~GeV). 
In the case of LS~5039, an ejection process, probably fed by an 
accretion disk, 
is clearly proved by a VLBA map which shows bipolar jets emerging from a 
central core. 
Its high L$_{\gamma}$ is tentatively explained by inverse Compton scattering. 
In \src, VLBI observations (at 5~GHz) show a possible jet-like elongation at 
milliarcseconds scales, suggesting the presence of a one-sided collimated radio jet, 
as found in several other X--ray binaries. 
A lower limit of 0.4~c for the intrinsic velocity of the radio jet has 
been derived (Massi et al. 2001).

We report here the results of a deep XMM-Newton observation performed in January 2005,
together with spectral analysis of archival \sax\ observations never reported in literature,
and for completeness we studied also 
other five short archival \xmm\ observations recently analysed by Chernyakova et al. (2006).

\section{Observations and data reduction}

\subsection{XMM-Newton observations}

The \xmm\ Observatory includes
three 1500~cm$^2$ X--ray telescopes each with an European Photon
Imaging Camera (EPIC) at the focus. Two of the EPIC imaging
spectrometers use MOS CCDs (Turner et al.~\cite{t:01}) and one
uses a pn CCD (Str\"uder et al. \cite{st:01}). 
Behind two of the
telescopes there are Reflection Grating Spectrometers (RGS,
0.35--2~keV; den Herder et al. \cite{dh:01}). 
\src\ was observed on 2005 January 27-28 for a net exposure of 48.7~ks.
Data were processed using version 6.5 of the \xmm\ Science Analysis
Software (SAS). Known hot, or flickering, pixels and electronic
noise were rejected using the SAS.
A further cleaning was necessary because of the presence of
a soft proton flare at the beginning of the observation, reducing the net good exposure times
to 42.6~ks and 30.3~ks, respectively for the EPIC MOS and the pn.
EPIC MOS used Imaging Prime Partial Mode, while EPIC pn the Small Window Mode.
All EPIC cameras used the medium
thickness filter.

Cleaned MOS (with pattern selection from 0 to 12)
and pn (patterns from 0 to 4) events files have been extracted,
and used for the subsequent analysis with {\sc xmmselect}.
The source spectra have been extracted from a circular
region with 40\arcsec\ radius, for the MOS1, MOS2 and pn
separately.
Background spectra have been always extracted from source free regions
of the same observation. When time selected spectra have been
considered, the same time selection has been applied also to the 
background.
Spectra have always been rebinned such that at least 20 counts per
bin were present and such that the energy resolution was not 
over-sampled by more than a factor 3.
The response matrices have been generated using the {\sc rmfgen}
 and {\sc arfgen} tools implemented in the SAS~6.5.
Free relative normalizations between MOS1, MOS2 and pn instruments were included in
the spectral fitting.
All spectral
uncertainties and upper-limits are given at 90\% confidence for
one interesting parameter.

We also analysed other 5 much 
shorter \xmm\  observations, performed in 2002.
Table~\ref{tab:log} gives a summary of the observations and
the corresponding orbital ($\theta_1$)
and  super-orbital phases ($\theta_2$).
The mean error on the orbital phase is ${\pm 0.03}$.
Here we adopt as time of the zero phase JD$_{0}$=2,443,366.775,
together with the  updated estimates of the orbital (P$_{1}$=26.4960$\pm{0.0028}$ days)
and of the super-orbital period (a periodic modulation of the phase and amplitude of the
radio outbursts, with a period, P$_{2}$, of 1667$\pm{8}$ days) reported in Gregory (2002).

\begin{table*}[ht!]
\caption{Summary of the \xmm\ observations reported here. 
``Exp'' is the nominal exposure time of a single MOS camera.$\theta_1$
and $\theta_2$ are the orbital  and
super-orbital phases (Gregory 2002).} 
\label{tab:log}
\begin{tabular}[c]{lllllc}
\hline\noalign{\smallskip}
OBS ID. & Start time     & End time                &  Exp. & $\theta_1$ & $\theta_2$ \\
      & (dy~mon~yr~hr:mn)     & (dy~mon~yr~hr:mn)  &  (ks) &           &            \\
\noalign{\smallskip\hrule\smallskip}
00    &  27 Jan 2005 17:41 &  28 Jan 2005 07:37  &  48.7  &    0.61       &    0.02   \\
01    &  10 Feb 2002 11:26 &  10 Feb 2002 12:50  &   6.4  &    0.76       &    0.37   \\
02    &  05 Feb 2002 06:24 &  05 Feb 2002 03:30  &   6.4  &    0.55       &    0.37   \\
03    &  21 Feb 2002 16:02 &  21 Feb 2002 17:26  &   7.5  &    0.18       &    0.38   \\
04    &  17 Feb 2002 04:40 &  17 Feb 2002 06:04  &   6.4  &    0.01       &    0.37   \\
05    &  16 Sep 2002 02:29 &  16 Sep 2002 04:09  &   6.4  &    0.97       &    0.50   \\
\noalign{\smallskip\hrule\smallskip}
\end{tabular}
\end{table*}

\subsection{BeppoSAX observations}

\src\ was observed twice with \sax,
on 1997 September 22 at 19:43 and on 
1997 September 26 at 14:55 (UTC), 
with an on-source time of 12~ks and 8.6~ks, respectively.
We report the results obtained with 
the Low-Energy Concentrator Spectrometer (LECS;
0.1--10~keV; Parmar et al. \cite{p:97}), the Medium-Energy Concentrator
Spectrometer (MECS; 1.8--10~keV; Boella et al. \cite{b:97}) 
and  the  Phoswich
Detection System (PDS; 15--300~keV; Frontera et al. \cite{f:97}).

Counts for the
spectral analysis have been extracted from circular regions with standard radii
(4$'$ for the  MECS and 8$'$ for the LECS).
The LECS and MECS spectra were rebinned to oversample the full
width half maximum of the energy resolution by
a factor 3 and to have  a minimum of 20 counts
per bin.
Background spectra were extracted from annular regions around the
source, in the same observations. 
Response matrices appropriate for the sizes of the extraction regions
were used.
In the spectral analysis we fitted together LECS (selected in the best calibrated 
energy range 0.1--4.0 keV) and MECS spectra (1.8--10~keV),
with free relative normalizations  between the instruments.

The PDS consists of four independent non-imaging
units arranged in pairs each having a
separate collimator. Each collimator was alternatively
rocked on- and 210\arcmin\ off-source every 96~s during
the observations.
After a proper cleaning of the events files aimed at excluding the ``spikes''
due to single particles hits, we extracted net spectra from the two 
observations.
Inspection and comparison of the background spectra show that there are no known 
contaminating sources in the background region. 
Since the only nearest possible high energy contaminating source is
QSO 0241+622, which is located just outside the PDS field of view, at 1.4$^\circ$,
we are confident that the high energy emission is produced by \src.

\section{Results}

\subsection{The XMM-Newton observations}

\begin{figure*}
  \centering
   \includegraphics[angle=-90,width=13.5cm]{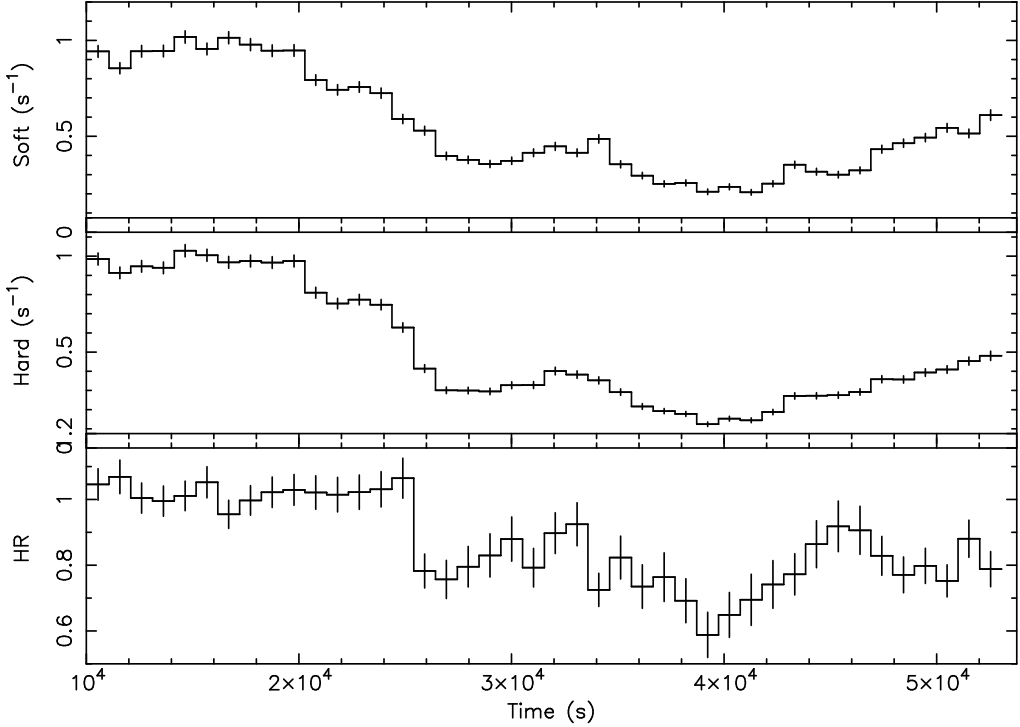}
      \caption{Background subtracted EPIC pn
lightcurves in two energy ranges (soft=0.3--2~keV; hard=2--12~keV)
together with their hardness ratio (HR=2--12~keV / 0.3--2~keV). 
Time is in seconds starting from 2005, January 27, at 17:56:26.
Bin time is 1024~s. }
         \label{fig:lc}
   \end{figure*}

\begin{figure*}
  \centering
   \includegraphics[angle=-90,width=13.5cm]{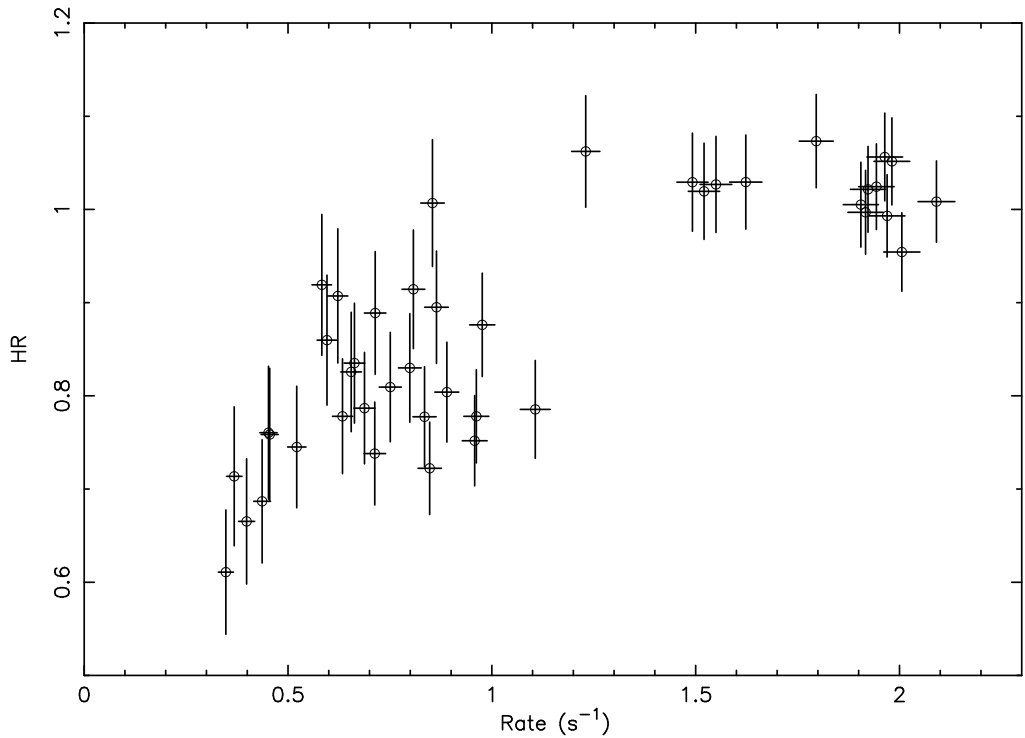}
      \caption{Hardness ratio (HR) versus Intensity (count rate in the total
energy range 0.3-12~keV) plot
for the longest \xmm\ observation (EPIC pn), with a bin time of 1024~s.
The hardness ratio is defined as  ratio between net counts
in the hard range 2--12~keV and  net counts in the soft energy range 0.3--2~keV.
}
         \label{fig:hr_intensity}
   \end{figure*}

The EPIC pn background subtracted lightcurves in two energy ranges (0.3--2~keV
and 2--12~keV) are reported in Fig.~\ref{fig:lc}, together with
their hardness ratio (HR). 
Both lightcurves show a smooth modulation,
ranging from a maximum rate around 1~counts~s$^{-1}$ down to about 0.2~counts~s$^{-1}$, in few hours.
The hardness ratio  shows a rather constant value 
of 1 for the first 15,000~s, and then 
a  sharp decrease, where it falls down to 0.8 in less than 1000 seconds.
After this decrease, the hardness ratio  is smoothly modulated, displaying
a wave-like shape, with two maxima (around 30,000~s and 46,000~s in Fig.~\ref{fig:lc})
and one minimum.
A plot of HR versus the source intensity 
(Fig.~\ref{fig:hr_intensity}) indicates that  the source is 
harder when it is brighter.

\vspace{2cm}

\vspace{2cm}

\begin{figure*}
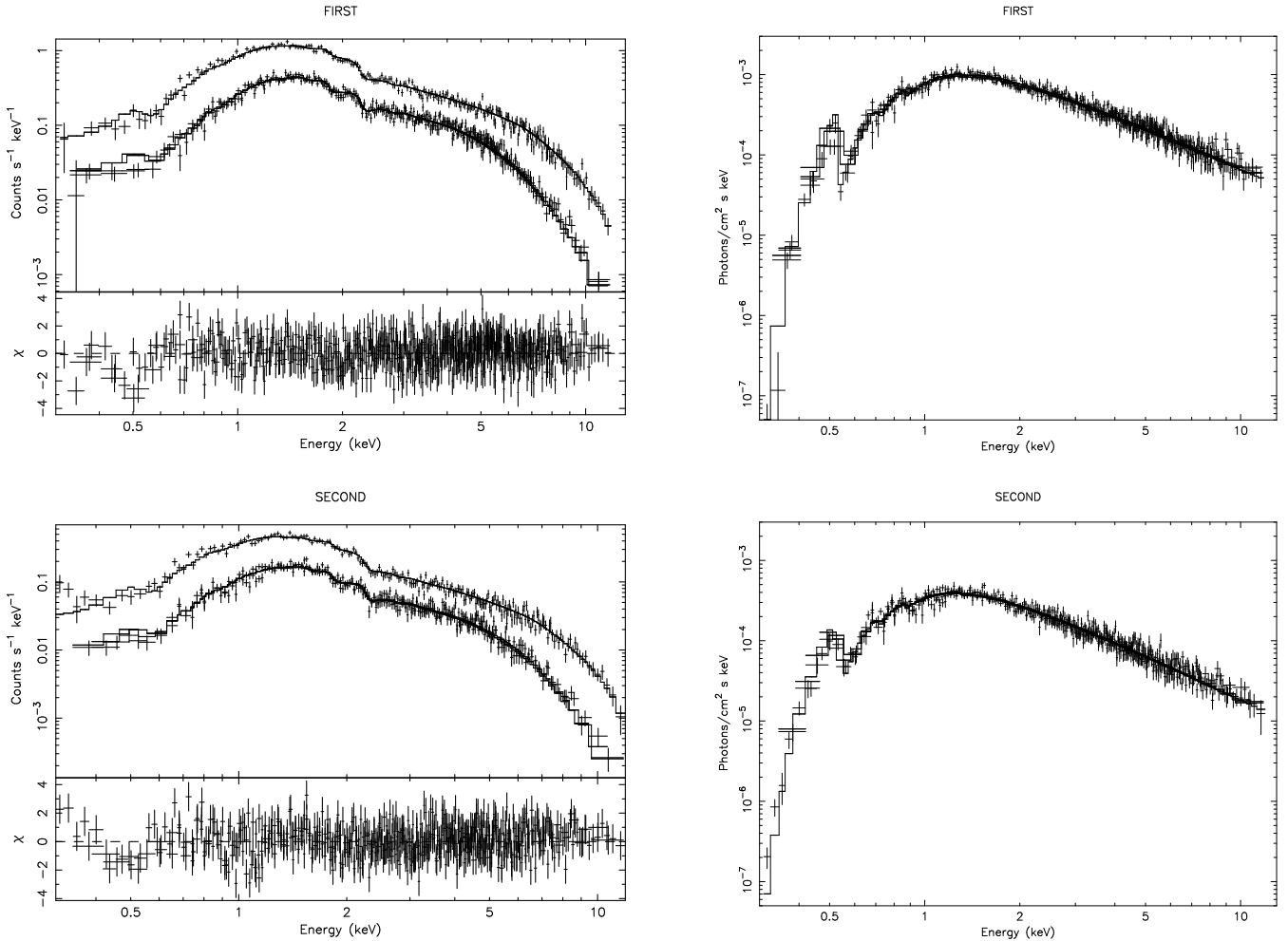

\hbox{\hspace{0.5cm}
\includegraphics[height=8.7cm,angle=-90]{5933fig3a.ps}
\hspace{1.0cm}
\includegraphics[height=7.9cm,angle=-90]{5933fig3b.ps}}
\vbox{\vspace{0.1cm}}
\hbox{\hspace{0.5cm}
\includegraphics[height=8.7cm,angle=-90]{5933fig3c.ps}
\hspace{1.0cm}
\includegraphics[height=7.9cm,angle=-90]{5933fig3d.ps}}
\caption[]{{\em Left panels:}  Two \src\  time selected 
counts spectra (MOS1+MOS2+pn), together with the residuals in units of standard deviations
when fitted with an absorbed powerlaw. From top to bottom, the {\em first} and the {\em second}
 spectra are shown.
{\em Right panels:}  The two corresponding time selected photon spectra. 
}
\label{fig:specall1}
\end{figure*}

Since HR is variable, 
we extracted two spectra, one corresponding to the first 14,000~s, when HR was 
constant and the second one for the remaining part of the observation.
In both cases, the continua appear featureless, and an absorbed power-law model gives
the best-fit. 
There is no evidence for a break up to 12~keV, and
a bremsstrahlung model
is unacceptable, with a reduced $\chi$$^2$ larger than 1.4.
An additional component (a blackbody) added to the power-law continuum
is not formally required  by the residuals, and can
contribute at most 1\% of the  flux observed in the energy range 0.5--10~keV.
Adopting a single absorbed power-law,  there is no evidence for column density variations
between the two time-selected spectra, thus we
fixed it at 5$\times$10$^{21}$~cm$^{-2}$.
The best-fit spectral results 
are reported in Table~\ref{tab:spec_all} and shown in Fig.\ref{fig:specall1}.

The RGS spectra extracted from the whole observation appear featureless,
with a count rate of 0.019~counts~s$^{-1}$ 
and 0.025~counts~s$^{-1}$ in the RGS1 and RGS2, respectively (0.3--2~keV).
The combined RGS1 plus RGS2 spectra are well fitted 
($\chi$$^2$/dof=110.7/127) with an absorbed powerlaw 
with an absorbing column density of (6$\pm{1})$$\times$10$^{21}$~cm$^{-2}$, and
a poorly constrained photon index of 1.9$\pm{0.6}$.
The statistics is too low to allow a time/hardness  selected
RGS spectral analysis, so we will not discuss them further.

A search for pulsations resulted in negative results.

We have reduced and analysed in a similar way also the five  shorter \xmm\ observations.
Within each observation there is no evidence for significant changes
in hardness ratio, so we extracted spectra from the whole exposures
(after cleaning for possible contamination from proton flares).
The spectral results obtained fitting the X--ray emission with an 
absorbed power-law (and absorbing column density
fixed at 5$\times$10$^{21}$~cm$^{-2}$), are reported in Table~\ref{tab:spec_all}.

\subsection{BeppoSAX observations}

\begin{figure*}
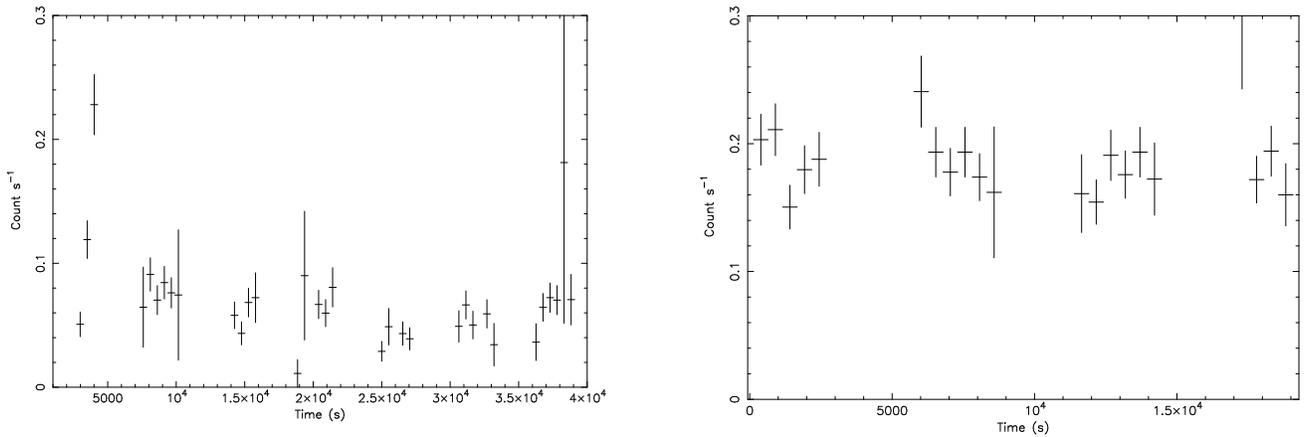

\hbox{\hspace{0.5cm}
\includegraphics[height=8.cm,angle=-90]{5933fig4a.ps}
\hspace{1.0cm}
\includegraphics[height=8.cm,angle=-90]{5933fig4b.ps}}
\caption[]{The BeppoSAX MECS2+MECS3 background subtracted lightcurves 
during the two observations (on the left and on the right, respectively).
The bin size is 512~s. A large flux variability is visible at the beginning of
the first \sax\ observation, with no evidence for simultaneous hardness ratio variations.
}
\label{fig:m23_lc}
\end{figure*}

The background subtracted lightcurves of \src\ during the two \sax\
observations are shown in Fig.~\ref{fig:m23_lc}.
The source was  on average $\sim$2.5 times brighter during the second
observation.
Within each  BeppoSAX observation the source displays some evidence
for flux variability, especially at the beginning of the first observation,
where a sort of ``flare'' is present, with 
the count rate increasing from $\sim$0.05~counts~s$^{-1}$ to 
$\sim$0.23~counts~s$^{-1}$ during the first
1500~s, without evidence, 
within the uncertainties, for a change in hardness ratio.

An absorbed power-law is the best-fit to  both
spectra (LECS+MECS2+MECS3), with the parameters reported
in Tab.~\ref{tab:spec_all}. We fixed the absorbing column density to
5$\times$10$^{21}$~cm$^{-2}$.
A bremsstrahlung model is not a good fit to the spectra, resulting
in a temperature higher than 18~keV, well above the MECS spectral range.
A single absorbed blackbody is unacceptable, with reduced $\chi$$^2$$>$2. 

The analysis of the PDS data 
resulted in a
detection at high energy (15--70 keV) only in the second
observation, which indeed catched the source at a higher flux level,
almost three times higher than in the first one.
The PDS count rates are 0.098$\pm{0.048}$~counts~s$^{-1}$ and 
0.31$\pm{0.07}$~counts~s$^{-1}$, respectively for the first and second observation (15--70~keV).
Thus, only the second can be considered a detection (at 4.4~$\sigma$).
This PDS rate translates into 
a 15--70 keV flux of about 3$\times$10$^{-11}$~erg~cm$^{-2}$~s$^{-1}$, 
adopting the best-fit reported in Table~\ref{tab:spec_all}.

\begin{table}[ht!]
\caption{Best-fit results of the overall spectral analysis.
$X00first$ and $X00second$ indicate the two time-selected 
spectra (MOS1+MOS2+pn, 0.3--12~keV) of the longest \xmm\ observation,
$X01-X05$ mark the five short \xmm\ observations, while $S01$ and $S02$
indicate the results from the two \sax\ pointings (LECS+MECS2+MECS3; 0.1--10 keV).
The best-fit model is and absorbed power-law, with
the absorbing column density fixed at 5$\times$10$^{21}$~cm$^{-2}$.
Fluxes are corrected for the absorption and are in units of 10$^{-12}$~erg~cm$^{-2}$~s$^{-1}$
in the energy range 2--10~keV. 
}
\label{tab:spec_all}
\begin{tabular}[c]{lllc}
\hline\noalign{\smallskip}
Spectrum      & Photon   &  Flux  &   $\chi$$^2$/dof \\
               &  Index             &  & \\
\noalign{\smallskip\hrule\smallskip}
X00first          &    1.62$\pm{0.01}$ & 12.9$\pm{0.1}$  & 680.7/596  \\ 
X00second         &    1.83$\pm{0.01}$ &  4.0$\pm{0.1}$  & 617.5/533  \\ 
X01       &        1.53$\pm{0.02}$ & 13.5$\pm{0.1}$           & 494.6/462 \\ 
X02       &        1.60$\pm{0.02}$ & 13.9$\pm{0.1}$           & 489.1/442 \\ 
X03       &        1.57$\pm{0.07}$ & 5.0$\pm{0.1}$            & 140.1/174 \\ 
X04       &        1.74$\pm{0.05}$ & 6.8$\pm{0.1}$            & 182.2/178 \\ 
X05       &        1.52$\pm{0.02}$ & 13.6$\pm{0.1}$           & 497.6/453 \\ 
S01       &        1.68$^{+0.12} _{-0.16}$ & 4.8$\pm{0.1}$    & 20.0/31 \\ 
S02       &        1.56$\pm{0.09}$ & 14.0$\pm{0.1}$           & 60.8/62 \\ 
\noalign{\smallskip\hrule\smallskip}
\end{tabular}
\end{table}

\section{Discussion and Conclusions}

\begin{figure*}
\hbox{\hspace{-0.2cm}
\includegraphics[height=6.5cm,angle=0]{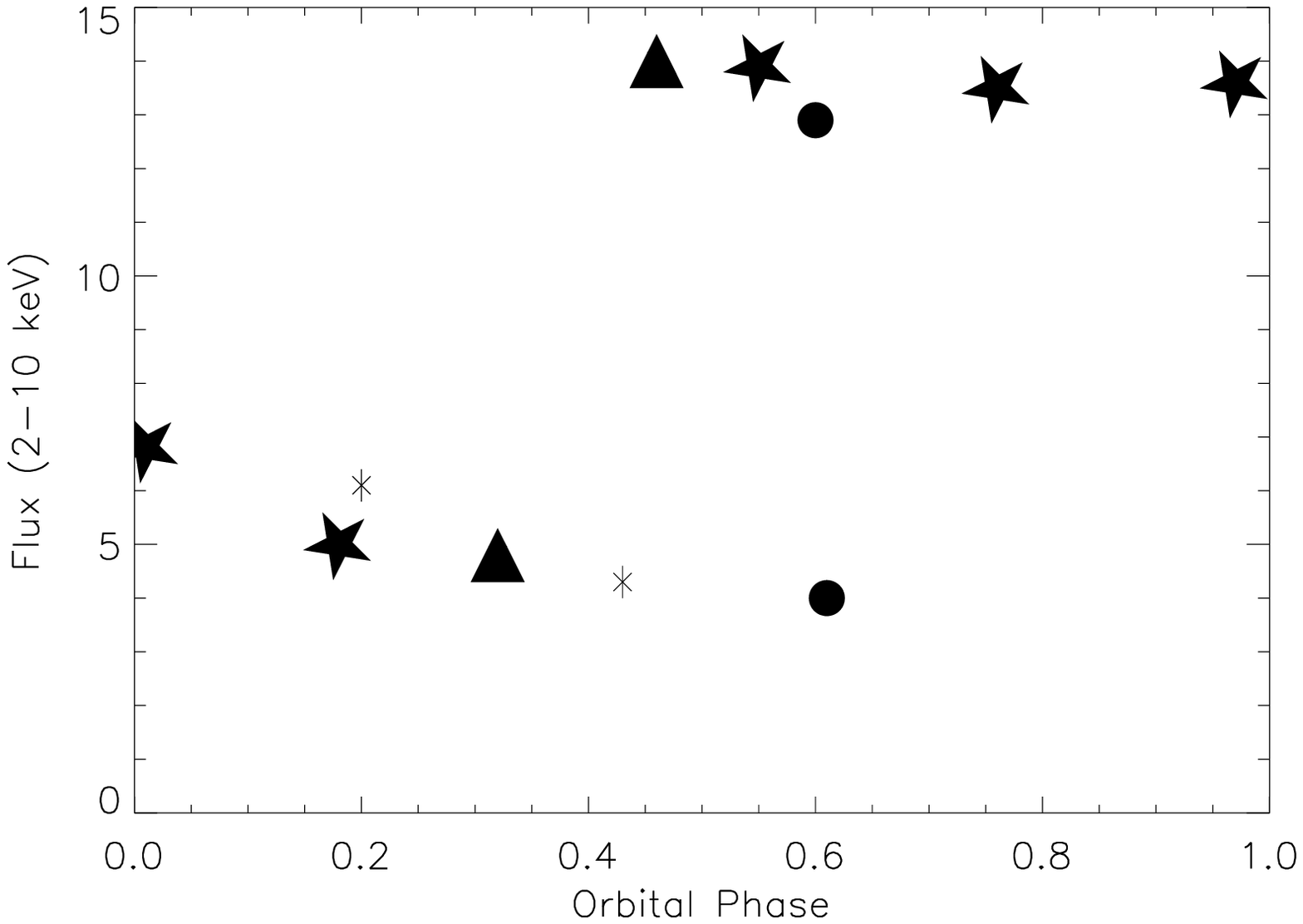}
\hspace{.0cm}
\includegraphics[height=6.5cm,angle=0]{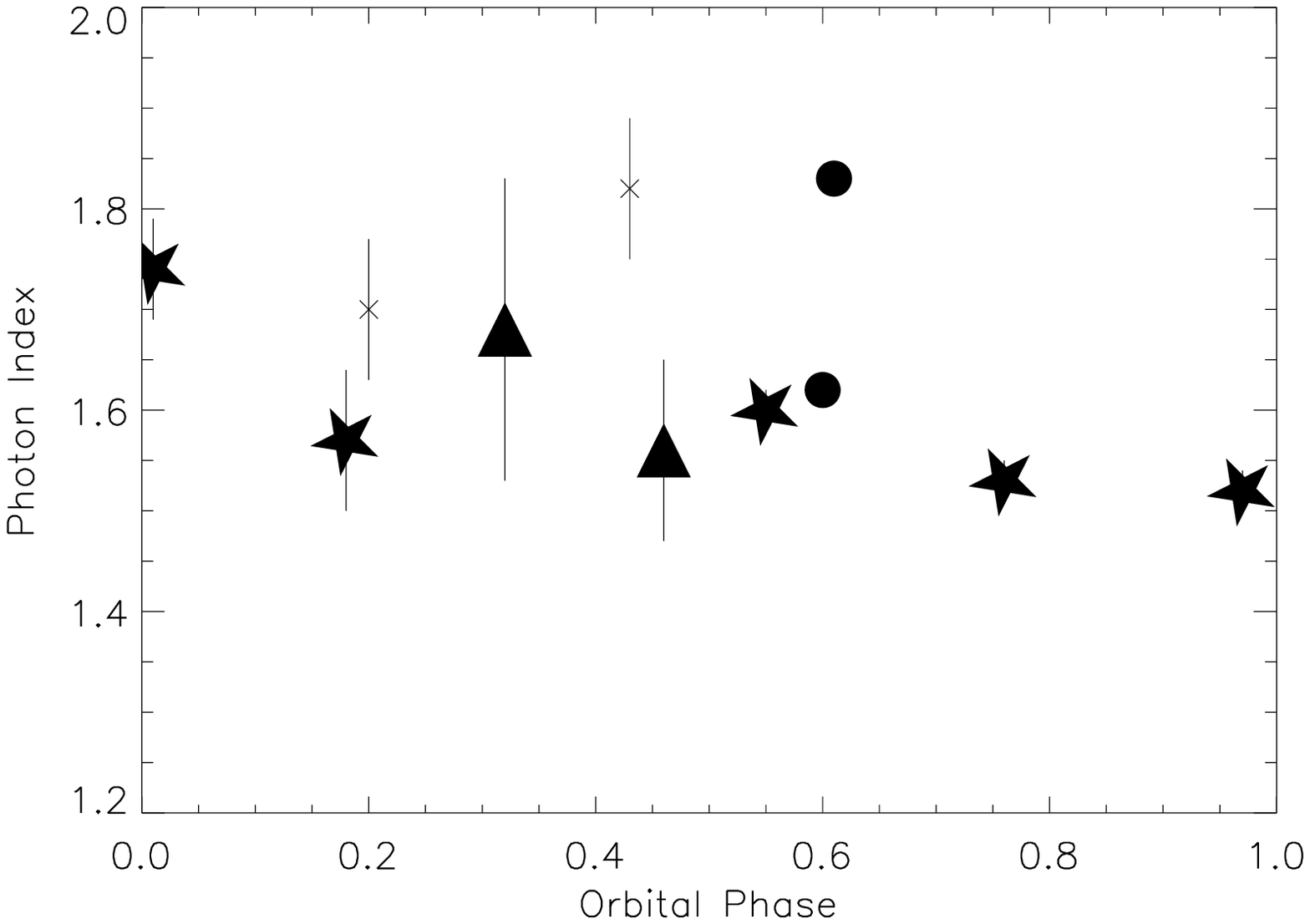}}
\vspace{-0.2cm}
\caption[]{Summary of the dependence with the orbital phase 
of the X--ray spectral parameters
of all the observations
analysed here (six \xmm\ and two \sax\ observations) together
with the ASCA results (published by Leahy et al. 1997) for completeness. 
The meaning of the symbols is as follows: {\em thick circles} represent
the longest \xmm\ observation, splitted into two (the ``first'' and the ``second''
spectra, with significantly different hardness ratio and flux);
the other 5 \xmm\ observations are marked with  {\em thick stars}, 
the 2 \sax\ observations are indicated with {\em thick triangles} and
the 2 ASCA observations (from Leahy et al. 1997)  with {\em thin crosses}. Fluxes
are corrected for the absorption and in units of 10$^{-12}$~erg~cm$^{-2}$~s$^{-1}$.
}
\label{fig:sum1}
\end{figure*}

\begin{figure*}
  \centering
   \includegraphics[angle=0,height=6.5cm]{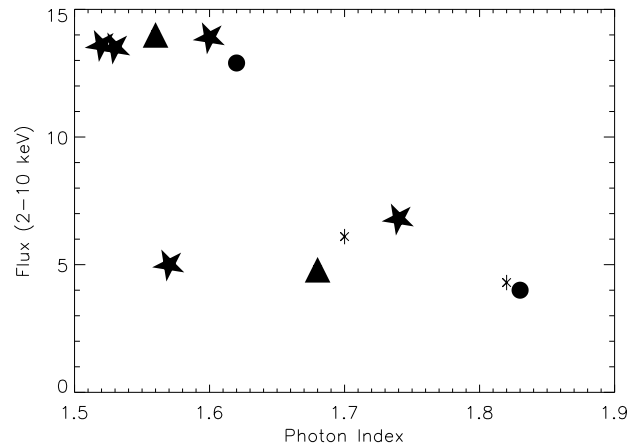}
      \caption{Unabsorbed flux (2--10 keV; 
in units of 10$^{-12}$~erg~cm$^{-2}$~s$^{-1}$) versus power-law photon index, collecting
all X--ray observations analysed here, together with the ASCA results 
(published by Leahy et al. 1997). 
The meaning of the symbols is the same as in Fig.~\ref{fig:sum1}.
The source appears to be softer when it is fainter. 
}
         \label{fig:sum2}
   \end{figure*}

We have reported here the deepest X--ray observation of \src\ ever carried out.
A systematic analysis of other five \xmm\ and two \sax\ observations has also been 
performed, in order
to investigate possible changes of the spectral parameters 
along the orbital phase (see Figs.~\ref{fig:sum1}
and ~\ref{fig:sum2}) and to get a global picture at X-rays.

During the longest \xmm\ observation for the first time we found  evidence for 
a rapid (on timescales of 
few hours) change in flux and hardness ratio: the source flux decreased by
a factor of $\sim$3, together with a drop in hardness ratio (see Fig.~\ref{fig:lc}) within
about 1000~s. 

This kind of variability (the source is harder when is brighter) is
also visible when comparing different observations performed in different times and 
with different satellites (Fig~\ref{fig:sum2}).
Another interesting variability feature is the presence of 
a sort of ``flaring'' behaviour at the beginning of one
of the \sax\ observations, but with no evidence for simultaneous 
hardness variations, mainly because of the low statistics.

The evolution of the source X--ray flux with the orbital phase
shows X--ray emission along all the phases, although with a difference of a factor
3 in the flux level between two states, with a ``high'' state preferentially
found in the phase range 0.4-1.0.
We found no obvious correlation of any of the 
X--ray spectral parameters with the superorbital period.

The analysis of archival \sax\ observations allowed the first
detection of the source with the PDS instrument (in the energy range 15--70 keV),
with evidence for a change in flux between a high and low state also in
the hard energy band.

The observations reported here demonstrate that the 
source is variable both at soft and hard X--rays, also at short timescales.
Thus, a detailed study of the overall spectrum should be in principle 
performed
with simultaneous observations at all wavelengths.
We try here to discuss the broad band energy spectrum distinguishing
at least two source states, a high and a low state, in order to get
informations on the source energetics.

\begin{figure*}
  \centering
   \includegraphics[angle=90,height=12.5cm]{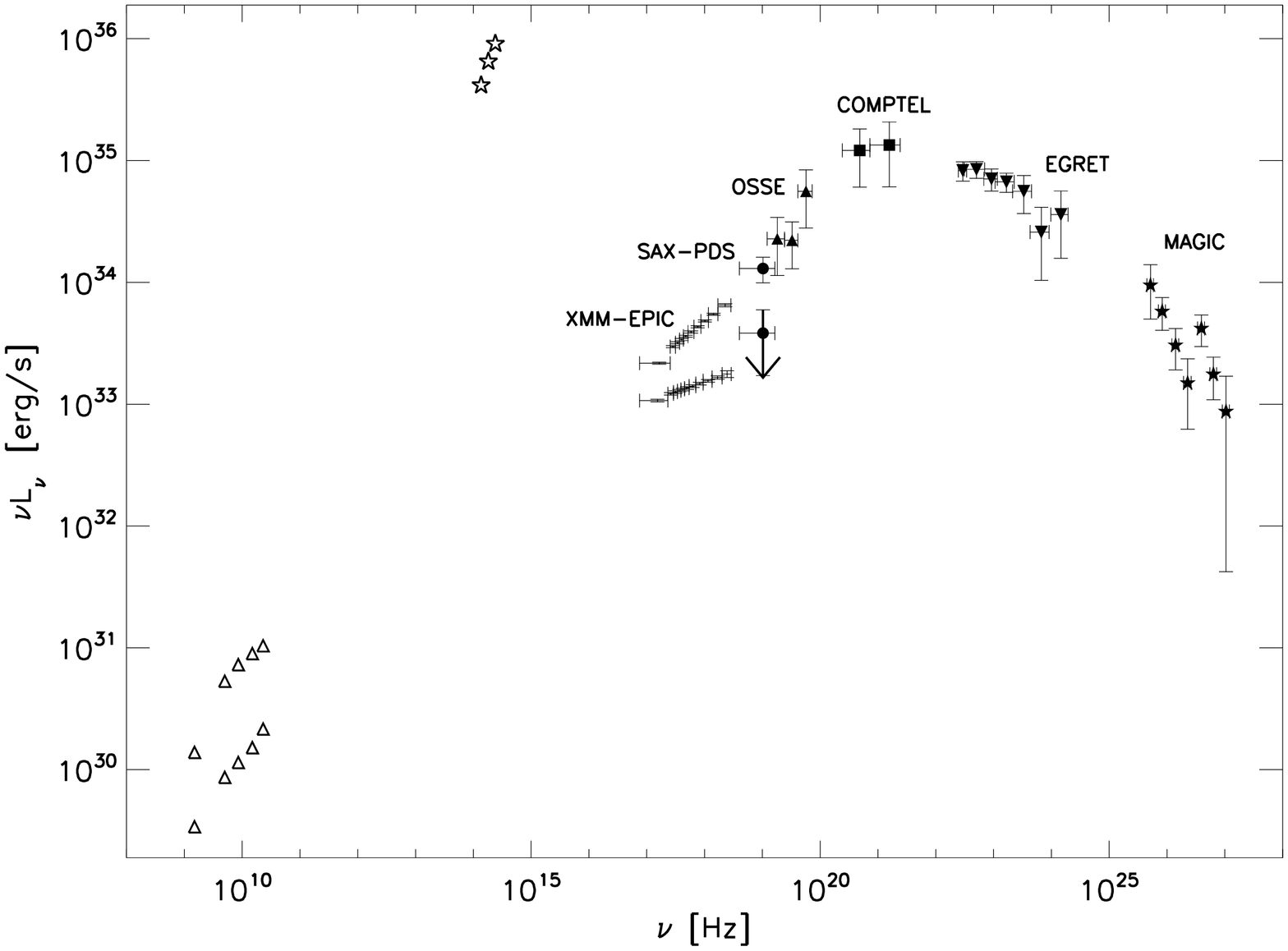}
      \caption{Broad band \src\ spectrum from radio wavelengths to TeV energies. 
Radio data for high and low states (open triangles), 
IR data (open stars) and OSSE data (filled triangles) are 
from Strickman et al. 1998. XMM-EPIC data (crosses) and SAX-PDS data 
(filled circles) both high and low states are from this work. 
COMPTEL data (filled squares) are from van Dijk et al. 1996. 
EGRET data (upside-down filled triangles) are from Kniffen et al. 1997. 
MAGIC data (filled stars) are from
Albert et al. 2006.
}
         \label{fig:sed}
   \end{figure*}

Figure~\ref{fig:sed} shows our \xmm\ and \sax\ results in the multiwavelength 
spectral energy distribution (SED) of \src.
For radio, \xmm\ and \sax\ data we show both high and low state spectra.
CGRO and MAGIC observations show low significance flux variations 
(peaking in the phase interval 0.4-0.7) correlated with those at X-ray energies. 
For these instruments we plot only average measurements 
of the datasets with positive detections.
During the high state, the keV-MeV spectrum can be fit with 
a hardening power law
($\Gamma$=1.6-1.5) up to the break in the COMPTEL energy range ($\sim$1-10 MeV). 
Then the spectrum shows a shallow roll-off  (EGRET: $\Gamma$=2.1; MAGIC: $\Gamma$=2.6) 
extending up to TeV energies.

Two main models to account for the SED of \src\ have been proposed.
The ``micro-quasar'' model involves streams of relativistic particles originating in episodes of 
super-Eddington accretion onto a compact star embedded in 
the mass outflow of the B-star
(Taylor \& Gregory 1984; Massi et al. 2004a, Massi 2004b). 
Alternatively \src\ might contain a non-accreting young pulsar 
in orbit around the mass-losing B-star (Maraschi \& Treves 1981; Dubus 2006).
In both cases, the high energy emission offers a diagnostics of the 
shock emission parameters in the jet or in the region where the 
hypothetical pulsar wind would interact with the circumstellar 
material.

The spectral index and break of the keV-MeV emission is consistent 
with that of diffusive shock acceleration models in the ``slow cooling" regime 
(i.e. far from energetic equilibrium between particle injection 
and cooling, see e.g. Chevalier 2000).
The shallow roll-off in the MeV-TeV energy range requires further 
model components in addition to the synchtrotron emission from 
shock-accelerated particles.
However, the large error bars of the observations in this 
energy range can hardly constrain the several possibile 
contributions based both on leptonic mechanisms as 
Compton-scattered stellar radiation, synchrotron self-Compton, 
pair cascades (Dubus 2006; Dermer \& B\"{o}ttcher 2006; 
Bosch-Ramon et al. 2006; Bednarek 2006), 
and on hadronic mechanisms as proton interactions, 
producing gamma-rays via neutral pion decay (Romero et al. 2003, 2005).

A clear signature of the SED is instead the spectral 
break in the 1-10 MeV range, possibly related to the maximum 
energy achievable from electrons in the shock acceleration 
accounting for synchrotron and Compton-scattered stellar radiation energy losses, 
the two major electron cooling mechanisms in the dense \src\ environment.
The maximum Lorentz factor of shocked particle is 
obtained by solving the equation:

\begin{equation}
\frac{1}{t_{\rm acc}}=\frac{1}{t_{\rm synch}}+\frac{1}{t_{\rm comp}}
\end{equation}
where $t_{\rm acc}=\xi t_{\rm gyr}=m_{\rm e} c \gamma \xi / e B $ is 
the electron acceleration time to reach a Lorentz factor $\gamma$ in 
a magnetic field with mean intensity $B$, and $\xi \goe 1$
because the acceleration time in Fermi processes cannot be smaller 
than the gyration time, i.e. an electron gains at most a fraction 
of its energy when executing a single gyration in first-order shock 
or second-order stochastic Fermi processes (see e.g. Dermer \& B\"{o}ttcher 2006).
$t_{\rm synch}=6 \pi m_{\rm e} c / \sigma_{\rm T} B^{2} \gamma $ is 
the synchrotron cooling time for an electron with Lorentz 
factor $\gamma$, and $t_{\rm comp}$ accounts for inverse Compton 
energy losses.
The solution of equation (1) is:

\begin{equation}
  \begin{array}{rll}
\gamma_{\rm max}=\sqrt{\frac{6 \pi e}{\sigma_{\rm T} \xi B (1+ {t_{\rm synch}}/{t_{\rm comp}})}} 
\\ \\
=\frac{1.2 \times 10^{8}}{\sqrt{\xi B[{\rm G}] (1+ {t_{\rm synch}}/{t_{\rm comp}})}}
  \end{array}
\end{equation}

If $t_{\rm synch} \ll t_{\rm comp}$ then the maximum synchrotron 
photon energy does not depend on the 
magnetic field B: $h \nu_{\rm max}=\hbar e B \delta \gamma_{\rm max}^{2} / m_{\rm e} c \simeq 167 \delta / \xi$ MeV, 
where $\delta$ is the Doppler factor to be included for jet scenarios ($1<\delta<2$, Chernyakova et al., 2006).
Thus, neglecting inverse Compton losses, the spectral break frequency would 
be more than one order of magnitude higher than that observed.

Since photons from the Be star have a mean 
energy of $2.7k_{\rm B}T \simeq 6.5$ eV ($T_{\rm s}=2.8 \times 10^{4}$ K), 
the turning point between Thompson and Klein-Nishina 
inverse Compton regime is for $\gamma_{\rm KN}=m_{\rm e}c^{2}/h \nu \simeq 8 \times 10^{4}$. 
Even in case of severe inverse Compton energy losses, eq. 2 shows that $\gamma_{\rm max}$ 
cannot be lower than $\gamma_{\rm KN}$ for typical magnetic fields values expected in 
microquasar jets or pulsar-stellar winds shocks (up to several Gauss). 
Thus $t_{\rm comp}=t_{\rm KN}$ and considering electrons with 
Lorentz factor $\gamma$ interacting with a blackbody surface radiation 
field (Blumenthal \& Gould 1970):
\begin{equation}
\begin{array}{rll}
t_{\rm KN}=\gamma \left(\frac{r}{R_{\rm s}}\right)^{2} \frac{64 \lambda_{\rm C}^{3}}{8 \pi^{3} c \sigma_{\rm T} \Delta^{2}} \left(\frac{m_{\rm e} c^{2}}{k_{\rm B}T_{\rm s}}\right)^{2} \left[\ln\left(0.552 \Delta \gamma \frac{k_{\rm B}T_{\rm s}}{m_{\rm e} c^{2}}\right)\right]^{-1}  
\\ \\
    \approx 10^{-30} \gamma r[cm]^{2} ~~{\rm s}
  \end{array}
\end{equation}
where $R_{\rm s}=13.4R_{\odot}$ is the Be star radius, $r$ 
is the distance between the star and the shock region and $\Delta\loe1$ accounts
for the Lorentz boosting of the stellar radiation field into 
the co-moving frame of the emission region (in the microquasar scenario).
 
Substituting eq. 3 in eq.2 we obtain a general expression for 
the maximum Lorentz factor of the electrons accounting both for 
synchrotron and inverse Compton losses in the KN regime:

\begin{equation}
\gamma_{\rm max} \simeq \sqrt{\frac{1.44 \times 10^{16} B[{\rm G}] - 8 \times 10^{38} \xi r[{\rm cm}]^{-2}}{\xi B[{\rm G}]^{2}}}
\end{equation}
which corresponds to a break energy of

\begin{equation}
h\nu_{\rm max}=\delta \left(\frac{167}{\xi}-\frac{10^{25}}{B[{\rm G}]r[{\rm cm}]^{2}}\right) ~~{\rm MeV}
\end{equation}

The observed break energy at 1-10 MeV constrains the 
value of $Br^{2} \approx 10^{23}$ G cm$^{2}$.
Shock models associated to pulsar wind interacting with 
the circumstellar environments predict a standoff distance 
of the order of $r \approx 10^{11}$ cm and magnetic fields of 
a few Gauss (Dubus 2006) in good agreement with the above constraint.

On the other hand, in accreting models, a shock originated in an 
extended jet ($r \approx 10^{14}-10^{15}$ cm; Massi et al. 2001, 2004a) 
would imply unrealistic low values for the magnetic fields, unless 
other cooling processes apart from inverse Compton on stellar 
radiation could play a major role (e.g. SSC, Gupta \& B\"{o}ttcher 2006). 
Shocks originating in the inner jet 
at a distance comparable with the compact object-star 
separation ($r \approx 10^{12}$ cm) would imply more realistic values for the magnetic fields.

Future deeper and simultaneous observations in the MeV--GeV 
energy range with AGILE and GLAST satellites and in VHE gamma-rays with 
MAGIC and HESS telescopes will better assess the IC parameters 
(and/or hadronic mechanisms) and thus the actual emission region properties.


\vspace{2cm}

\vspace{2cm}


\begin{acknowledgements}
The \xmm\ data analysis is supported by the Italian Space Agency (ASI),
through contract ASI/INAF I/023/05/0.
\end{acknowledgements}


\begin{thebibliography}{}

\bibitem[]{}Albert, J., Aliu, E., Anderhub, H., et al., 2006, Science, 312, 1771

\bibitem[]{}Apparao, K.M.V., 2001, A\&A, 366, 865 

\bibitem[]{}Aharonian, F., Akhperjanian, A.G., Aye, K.M., et al., 2005, Science, 309, 746

\bibitem[]{}Bednarek, W., MNRAS in press (astro-ph/0606421)

\bibitem[]{}Bignami, G.F.B., et al., 1981, ApJ Letters, 247, L85. 

\bibitem[]{}Blumenthal, G. R. \& Gould, R. J., 1970, Reviews of Modern Physics, 42, 237

\bibitem[1997]{b:97}Boella, G., Chiappetti, L., Conti, G., et al., 1997, A\&AS, 122, 327

\bibitem[]{}Bosch-Ramon, V., Romero, G. E., Paredes, J. M., 2006, A\&A, 447, 263

\bibitem[]{}Campana, S., Stella, L., Mereghetti, S., Colpi, M., 1995, 297, 385 

\bibitem[]{}Chernyakova, M., Neronov, A., Walter, R., 2006, MNRAS in press (astro-ph/0606070)

\bibitem[]{}Chevalier, R. A., 2000, ApJ, 539, L45

\bibitem[2001]{dh:01}Den Herder, J. W., Brinkman, A. C., Kahn, S. M., et al. 2001, A\&A, 365, L7

\bibitem[]{}Dermer, C. D. \& B\"{o}ttcher, M., 2006, ApJ, 643, 1081

\bibitem[]{}Dickey, Lockman, 1990, ARAA 28, 215

\bibitem[]{}Dubus, G., 2006, A\&A in press (astro-ph/0605287)

\bibitem[]{}Fender R., 2001, in ``Relativistic flows in Astrophysics", Springer Verlag 
Lecture Notes in Physics, Eds A.W. Guthmann, M. Georganopoulos, K. Manolakou and A. Marcowith

\bibitem[]{}Frail, D.A., Hjellming, R.M., 1991, AJ, 101, 2126 

\bibitem[1997]{f:97}Frontera F., Costa E., Dal Fiume D., et al., 1997, A\&AS 122, 371

\bibitem[]{}Goldoni, P., Mereghetti, S., 1995, A\&A, 299, 751 

\bibitem[]{}Gregory, P.C., Taylor, A.R., 1978, Nature, 272, 704 

\bibitem[]{}Gregory, P.C., Peracaula, M., Taylor, A.R., 1999, ApJ, 520, 376 

\bibitem[]{}Gregory, P.C., 2002, ApJ, 575, 427 

\bibitem[]{}Greiner, J., Rau, A., 2001, A\&A 375, 145 

\bibitem[]{}Gupta, S. \& B\"{o}ttcher, M., 2006, submitted to ApJ (astro-ph/0606590)

\bibitem[]{}Harrison, F.A., Ray, P.S., Leahy, D.A., et al. 2001, ApJ 528, 454 

\bibitem[]{}Hermsen, W., et al., 1977, Nature, 269, 494. 

\bibitem[]{}Hutchings, J.B., Crampton, D., 1981, PASP, 93, 486 

\bibitem[]{}Kniffen, D.A., Alberts, W.C.K., Bertsch, D.L., et al., 1997, ApJ 486, 126 

\bibitem[]{}Leahy, D.A., Harrison, F.A., Yoshida, A., 1997, ApJ, 475, 823 

\bibitem[]{}Maraschi, L., Treves, A. 1981, MNRAS, 194, 18 

\bibitem[]{}Massi, M., Rib\'o, M., Paredes, et al., 2001, A\&A, 376, 217

\bibitem[]{}Massi, M., Rib\'o, M., Paredes, et al., 2004a, A\&A, 414, L1

\bibitem[]{}Massi, M., 2004b, A\&A, 422, 267

\bibitem[]{}Mendelson, H., Mazeh, T., 1989, MNRAS, 239, 733 

\bibitem[]{}Paredes, J.M., Estallela, R., Ruis, A. 1990, A\&A, 232, 337 

\bibitem[]{}Paredes, J.M., Marziani, P., Mart\'i, J., et al. 1994, A\&A, 288, 519 

\bibitem[]{}Paredes, J.M., Mart\'i, J., Peracaula, M., Rib\'o, M. 1997, A\&A, 320, L25 

\bibitem[]{}Paredes, J.M., Mart\'i, J., Rib\'o, M.,  Massi, M., 2000, Science, 288, 2340 

\bibitem[]{}Paredes, J.M., Bosch-Ramon, V., Romero, G.E.,  2005, A\&A, 451, 259

\bibitem[1997]{p:97}
Parmar, A.N., Martin, D.D.E., Bavdaz, M., et al., 1997, A\&AS 122, 309

\bibitem[]{}Romero, G. E., Torres, D. F., Kaufman Bernad\'{o}, M. M., Mirabel, I. F., 2003, A\&A, 410, L1

\bibitem[]{}Romero, G. E., Christiansen, H. R., Orellana, M., 2005. ApJ, 632, 1093

\bibitem[]{}Strickman, M. S., Tavani, M., Coe, M. J., et al. 1998, ApJ, 497, 419

\bibitem[2001]{st:01}Str\"uder, L., Briel, U., Dennerl, K., et al. 2001, A\&A, 365, L18

\bibitem[]{}Taylor, A.R., Gregory, P.C., 1982, ApJ, 255, 210 

\bibitem[]{}Taylor, A.R., Gregory, P.C., 1984, ApJ, 283, 273 

\bibitem[]{}Taylor, A.R., Kenny, H.T., Spencer, R.E., Tzioumis, A., 1992, ApJ, 395, 268 

\bibitem[]{}Taylor, A.R., Young, G., Peracaula, M., et al., 1996, A\&A, 305, 817 

\bibitem[2001]{t:01}Turner, M. J. L., Abbey, A., Arnaud, M., et al. 2001, A\&A, 365, L27

\bibitem[]{}van Dijk, R., Bennet, K., Bloemen H., et al. 1996. A\&A, 315, 485

\bibitem[]{}Zamanov, R.K., Mart\'i, J., Paredes, J.M., et al. 1999, A\&A, 351, 543

\end{thebibliography}
\end{document}